\documentclass[useAMS, usenatbib, onecolumn]{mn2e}
\usepackage{graphicx}
\usepackage{subfigure}

\newcommand{\vc}[1]{\textit{\textbf{#1}}}
\newcommand{\alphen}[1]{
  \renewcommand{\theenumi}{(\alph{enumi})}
  \setcounter{enumi}{#1}}

\title[A partially dimensionally-split approach to numerical MHD]{A partially dimensionally-split approach to numerical MHD}
\author[H. Sriskantha, M. Ruffert]{H. Sriskantha\thanks{E-mail: hari.sriskantha@ed.ac.uk}, M. Ruffert\\
Maxwell Institute for Mathematical Sciences, School of Mathematics, The University of Edinburgh, Edinburgh, EH9 3JZ, UK.}

\usepackage{times}

\begin{document}

\date{Completed March 2013.}
\pagerange{\pageref{firstpage}--\pageref{lastpage}} \pubyear{2013}
\maketitle

\label{firstpage}

\begin{abstract}
We modify an existing magnetohydrodynamics algorithm to make it more compatible with a dimensionally-split (DS) framework. It is based on the standard reconstruct-solve-average strategy (using a Riemann solver), and relies on constrained transport to ensure that the magnetic field remains divergence-free ($\nabla \cdot \vc{B} = 0$). The DS approach, combined with the use of a single, cell-centred grid (for both the fluid quantities and the magnetic field), means that the algorithm can be easily added to existing DS hydrodynamics codes. This makes it particularly useful for mature astrophysical codes, which often model more complicated physical effects on top of an underlying DS hydrodynamics engine, and therefore cannot be restructured easily. Several test problems have been included to demonstrate the accuracy of the algorithm, and illustrative source code has been made freely available online. 
\end{abstract}


\section{Introduction}
When developing complex astrophysical codes, we often start with a simple hydrodynamics engine. This is designed to numerically solve the Euler equations of pure hydrodynamics, which -- in conservative form -- can be written as: 
\begin{equation}
	\frac{\partial\vc{U}}{\partial t}  +  \frac{\partial \vc{F}}{\partial x} + \frac{\partial\vc{G}}{\partial y}  + 		\frac{\partial\vc{H}}{\partial z}  = 0,
	\label{eqn:conservationlaw}
\end{equation}
where $\vc{U}$ is the vector of state quantities we wish to evolve in time, and the flux vectors $\vc{F}(\vc{U})$, $\vc{G}(\vc{U})$ and $\vc{H}(\vc{U})$ contain their respective fluxes in the three spatial (Cartesian) dimensions. A popular approach for solving these equations numerically is the reconstruct-solve-average (RSA) strategy \citep[a term coined by][]{MIG07}. We start by dividing our space into a 3D grids of `cells', and then we discretise the initial data by storing only the cell-volume averages of the state quantities at every cell centre. Each timestep then comprises three stages. First, we \textit{reconstruct} this discretised data to estimate the value of the state quantities on either side of each cell face, thus defining -- in effect -- a series of `Riemann problems'.\footnote{A Riemann problem is defined as the solution to a conservation law with piecewise constant initial data and a single discontinuity \citep{LEV92}. Algorithms exist to find this solution both exactly and approximately (the latter sacrificing some accuracy in favour of computational speed).} Next, we \textit{solve} each of these problems independently using a Riemann solver algorithm, which determines the fluxes of the state quantities through each cell face. Finally, we use these inter-cell fluxes to determine the new cell-averaged state quantities. These three stages are repeated until the desired amount of `time' has been simulated.

This strategy is often combined with dimensional splitting (DS), where the 3D grid is split into several independent 1D problems and each timestep then becomes a series of `directional passes'. In each of these passes, all of the rows in a given direction are evolved according to the RSA strategy above. However, we only deal with the cell faces which are perpendicular to the direction of the pass: the $x$-direction pass determines the flux vector $\vc{F}(\vc{U})$ through all the $x =$ constant faces, while the $y$- and $z$- direction passes determine $\vc{G}(\vc{U})$ and $\vc{H}(\vc{U})$ through the $y =$ constant and $z =$ constant faces respectively. The popularity of this approach is due to its many advantages: the timestep is not as constricted by stability conditions as with a fully multi-dimensional code; it is simple to upgrade an existing 1D code to more dimensions in this way; the rows in each pass can be evolved independently, which means the code can be easily parallelised; it has been shown that the accuracy of the 1D code can be preserved by alternating the order of the passes \citep{LIS03}.

Once this hydrodynamics engine is in place, it can be improved by adding other physical phenomena (such as viscosity, self-gravity or a realistic equation of state), and more sophisticated computational features (such as parallelisation and adaptive/non-uniform grids). This is generally an effective way to develop code, but can cause difficulties if we then need to modify the underlying engine -- for example, when adding the effects of magnetic fields. The dynamics of a fluid in a magnetic field are governed by the equations of magnetohydrodynamics (MHD), which combine the Euler equations with Maxwell's equations of electromagnetics. Although these equations can be written in the same form as eq. (\ref{eqn:conservationlaw}), they cannot be easily incorporated into DS-RSA codes because of the need to maintain the divergence-free condition:
\begin{equation}
    \nabla \cdot \vc{B} = 0.
     \label{eqn:divergencefreecondition}
\end{equation}
Ignoring this condition in numerical simulations can result in unwanted effects, such as the creation of unphysical forces parallel to the magnetic field, and the loss of conservation of momentum and energy \citep{BAL99}. 

There are several families of algorithms for maintaining this condition \citep[which are summarised in e.g.][]{TOT00}. In this paper, we focus on the constrained transport (CT) family, where the magnetic field quantities are discretised in such a way that $\nabla \cdot \vc{B}$ does not change in time analytically (to machine error). We then simply choose initial conditions such that $\nabla \cdot \vc{B} = 0$. Many CT algorithms have been developed \citep[such as][]{DAI98, BAL99, TOT00, LON00}, but these are often dimensionally unsplit and/or require a `staggered' grid: the fluid quantities are stored as volume-averages at cell centres, while the magnetic field quantities are stored as \textit{surface}-averages at cell \textit{faces}. This is not troublesome for researchers intending to develop new codes from scratch, but is less than ideal for those working on mature codes as described above. Even implementing the non-staggered algorithm suggested in \citet{TOT00} -- which was modified from the algorithm of \citet{BAL99} -- would require considerable restructuring of the underlying hydrodynamics engine. Instead, a preferable algorithm is one that minimises the need for such restructuring. In this paper, we show that one can be developed by reframing T\'oth's approach so that it is more compatible with a dimensionally-split framework. By doing so, we aim to considerably simplify the development required to add the MHD equations to mature astrophysical codes.

In \S2 we describe some practical details, including notation, units and the formulation of the MHD equations used. In \S3 we summarise the Balsara-Spicer-T\'oth algorithm for maintaining the divergence-free condition, while in \S4 we derive our `more' dimensionally-split version. Readers who are only interested in the final algorithm may skip ahead to \S5, where we include a clear step-by-step guide to implementing it. (Additionally, illustrative source code has been made available online -- which was designed to be as easy as possible to be read -- in order to aid understanding of the algorithm.) Finally, in \S6 we demonstrate its accuracy with several well-known MHD tests.

\section{Formulation and Notation}

\subsection{Discretisation}
\label{sec:discretisation}
\begin{figure}
	\centering
	\includegraphics[width=\linewidth]{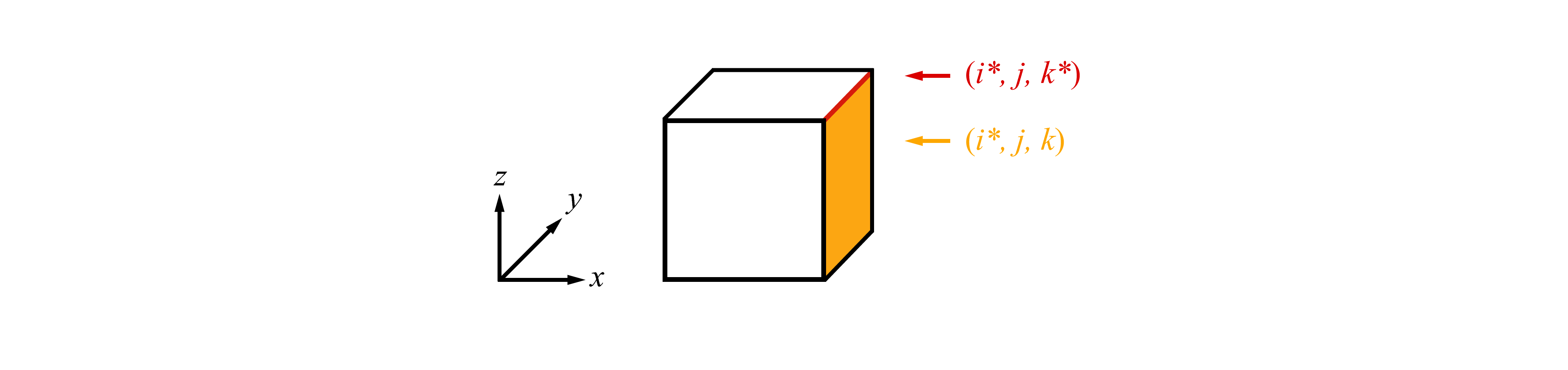}
	\caption{This figure depicts the cell centred at position $(i, j, k)$. The orange face is labelled $(i*, j, k)$, where $i* = i + 1/2$, because it is located $\frac{1}{2} \Delta x$ away from the cell centre. Using similar reasoning, the red edge is labelled $(i*, j, k*)$. (See \S\ref{sec:discretisation} for details.)}
	\label{img:cell}
\end{figure}

We model Cartesian space in the domain $D_x \times D_y \times D_z$. We divide this space into $n_x \times n_y \times n_z$ `cells' in the usual way, each of volume $\Delta x \times \Delta y \times \Delta z$, such that $D_x = n_x \cdot \Delta x$, etc. (We assume that the cell sizes are constant and uniform, but it is simple to extend our algorithm to adaptive/non-uniform grids.) The position of each cell centre is labelled as $(i, j, k)$, where  $i \in \{1, 2, ... ,n_x\}$, etc. We also refer to the faces and edges of cells, using starred notation $i* = i + 1/2$. A face will therefore have one starred co-ordinate, e.g. $(i*, j, k)$, while an edge will have two, e.g. $(i*, j*, k)$. A typical cell, face and edge are depicted in Fig. \ref{img:cell}. Finally, we discretise time such that $t_n = n \cdot \Delta t_n$. The timestep $\Delta t_n$ can change with time, but must always satisfy the Courant-Friedrich-Lewy (CFL) stability condition. 

To avoid excessive indices, we only write down spatial co-ordinates that \textit{differ} from $(i, j, k)$. For example, the face-average $\vc U^{i*, j, k+1}_n$ will be abbrievated to $\vc U^{i*, k+1}_n$. The reader should therefore assume that there are always three spatial co-ordinates, even if all three are not explicitly written down. When there is no change from $(i, j, k)$ we use $^\bullet$, such that $\vc U^{i, j, k}_{n}$ is abbrievated to $\vc U^{\bullet}_{n}$. Similarly, we do not use subscripts to denote components of discretised values. For example: the cell average of the $x$-component of the magnetic field will be written as $(Bx)^\bullet_n$, not $B^\bullet_{x, n}$. A summary of the non-standard notation used throughout this paper is provided in Table \ref{tbl:notationsummary}.

\begin{table}
 	\caption{A summary of the non-standard notation used throughout this paper. See \S\ref{sec:discretisation} for more details.}
 	\label{tbl:notationsummary}
 	 \begin{tabular}{ll}
	\hline
	Notation & Meaning \\
	\hline
	$^\bullet$ & the cell centred at $(i, j, k)$  \\
	$(j + 1)$ & the cell centred at $(i, j + 1, k)$ \\
	$i*$ & the index $i + 1/2$ \\
	$(i*, k+1)$ & the cell face centred at $(i*, j, k+1)$ \\
	$\tilde{\vc{F}}$, $\tilde{\vc{G}}$, $\tilde{\vc{H}}$ &  inter-cell flux vectors returned by the Riemann solver \\
	$(\tilde{F}7)$ & the 7th component of $\tilde{\vc{F}}$ \\
	$(Bx)$& the $x$-component of $\vc{B}$ \\
	$\varphi$ & $\varphi =1$ or $1/\sqrt{4\pi}$ depending on units (see \S\ref{sec:units})\\
	\hline
 	 \end{tabular}
\end{table}

Using the discretisation above, we can now write a numerical equation for the DS-RSA strategy, which gives $\vc{U}^\bullet_{n+1}$ in terms of $\vc{U}^\bullet_n$:
\begin{equation}
	\nonumber \vc{U}^\bullet_{n+1} = \vc{U}^\bullet_{n} - \frac{\Delta t}{\Delta x} (\tilde{\vc{F}}^{i*}_{n*} - \tilde{\vc{F}}^{i*-1}_{n*}) \label{eqn:ds-rsa-x} - \frac{\Delta t}{\Delta y} (\tilde{\vc{G}}^{j*}_{n*} - \tilde{\vc{G}}^{j*-1}_{n*}) \label{eqn:ds-rsa-y} - \frac{\Delta t}{\Delta z} (\tilde{\vc{H}}^{k*}_{n*} - \tilde{\vc{H}}^{k*-1}_{n*}) \label{eqn:ds-rsa-z},
\label{eqn:rsasteps}
\end{equation}
where the last three terms correspond respectively to the three directional passes. The terms $\tilde{\vc{F}}^{i*}_{n*}$, $\tilde{\vc{G}}^{j*}_{n*}$ and $\tilde{\vc{H}}^{k*}_{n*}$ are the inter-cell flux vectors returned by the Riemann solver during the `solve' stage (signified by tildes), and $n* = n + 1/2$ refers to the half-timestep.

\subsection{Units}
\label{sec:units}
In the literature, there are two sets of units commonly used for the magnetic field, which differ by a factor of $\sqrt{4\pi}$. To aid compatibility, we use unit-neutral notation, using a constant $\varphi$. For SI units, $\varphi =1$, while for CGS units, $\varphi = 1/\sqrt{4\pi}$. By including this constant in the code, we can switch easily between the two. We will make it clear which version is used by any referenced papers.

\subsection{The MHD equations}
The MHD equations can be written in the same form as eq. (\ref{eqn:conservationlaw}). The vector of state quantities is defined as:
\begin{equation}
  \vc{U} = [\rho, \rho v_x, \rho v_y, \rho v_z, \mathcal{E}, B_x, B_y, B_z]^T,
  \label{eqn:statevector}
\end{equation}
where $\rho$ is the mass density, $\rho \vc{v}$ is the momentum density, $\mathcal{E}$ is the total energy density, and $\vc{B}$ is the magnetic field density. Recall that the cell-averaged values of these quantities are stored at every cell centre. We refer to the first five elements as `fluid quantities' and the last three elements as `magnetic field quantities' whenever we need to distinguish between the two. The flux vectors are defined as:
\begin{equation}
 \vc{F}=
\left[\begin{array}{c}
         \rho v_x \\
	\rho v_x^2 + p_{\rmn{T}} - \varphi^2 B_x^2 \\
	\rho v_x  v_y - \varphi^2 B_xB_y\\
	\rho v_x v_z - \varphi^2 B_xB_z\\
	v_x (\mathcal{E} + p_{\rmn{T}}) - \varphi^2 B_x(\vc{v}\cdot\vc{B})\\
	0 \\
	-E_z \\
	E_y
        \end{array}\right],
\quad \vc{G} = 
\left[\begin{array}{c}
         \rho v_y \\
	\rho v_x u_y - \varphi^2 B_xB_y \\
	\rho v_y^2 + p_{\rmn{T}} - \varphi^2 B_y^2 \\
	\rho v_y v_z - \varphi^2 B_yB_z\\
	v_y (\mathcal{E} + p_{\rmn{T}}) - \varphi^2 B_y(\vc{v}\cdot\vc{B})\\
	E_z \\
	0 \\
	- E_x 
        \end{array}\right],
\quad \vc{H}= 
\left[\begin{array}{c}
         \rho v_z \\
	\rho v_x v_z - \varphi^2 B_xB_z\\
	\rho v_y v_z - \varphi^2 B_yB_z\\
	\rho v_z^2 + p_{\rmn{T}} - \varphi^2 B_z^2 \\
	v_z ( \mathcal{E} + p_{\rmn{T}}) - \varphi^2 B_z(\vc{v}\cdot\vc{B}) \\
	- E_y \\
	E_x \\	
	0 
        \end{array}\right],
\label{eqn:fluxvectors}
\end{equation}
where $p_{\rmn{T}}$ is the total pressure, such that $p_{\rmn{T}} = p_{\rmn{M}} + p_{\rmn{G}}$. The magnetic pressure is given by $p_{\rmn{M}} = \frac{1}{2}\varphi^2 B^2$, where $B^2 = \vc{B}\cdot\vc{B}$. The gas pressure, $p_{\rmn{G}}$, is related to the total energy density by an ideal equation of state, which closes the system, such that: 
\begin{equation}
  \mathcal{E} = \frac{p_{\rmn{G}}}{\gamma - 1} + \frac{1}{2}\rho \vc{v}^2 + \frac{1}{2}\varphi^2 {B}^2,
\label{eqn:equationofstate}
\end{equation}
where $\gamma$ is the ratio of specific heats. Finally, $\vc{E}$ is the electric field, which is given by the ideal version of Ohm's law: 
\begin{equation}
 \vc{E} = -\vc{v} \times \vc{B}.
 \label{eqn:ohmslaw}
\end{equation}

\section{Balsara-Spicer-T\'oth Algorithm}
We now describe the algorithm of \citet{BAL99}, later modified by \citet{TOT00}. The fluid quantities are evolved using the usual DS-RSA strategy, as described in \S1, while the magnetic field quantities are evolved using constrained transport.

\subsection{Constrained transport (CT)}
CT was first developed by \citet{EVA88}, and uses Faraday's law to determine how the magnetic field $\vc{B}$ changes in time:
\begin{equation}
  \frac{\partial}{\partial t}\vc{B} = - \nabla \times \vc{E}.
  \label{eqn:faradayslaw}
\end{equation}
where $\vc{E}$ is the electric field. If take the divergence of both sides of this equation, then we get:
\begin{equation}
   \frac{\partial}{\partial t}\nabla \cdot \vc{B} = - \nabla \cdot (\nabla \times \vc{E}) = 0.
\end{equation}
where the second equality is true because $\nabla \cdot (\nabla \times \vc{F}) = 0$ for any vector $\vc{F}$. This means that if we evolve the magnetic field using eq. (\ref{eqn:faradayslaw}), then $\nabla \cdot \vc{B}$ will not change in time analytically. We can therefore maintain the divergence-free condition by simply choosing initial conditions such that $\nabla \cdot \vc{B} = 0$  \citep{TOT00}. To implement eq. (\ref{eqn:faradayslaw}) numerically, we start by considering the $x$-component of the magnetic field averaged on the face $(i*)$, which we call $(bx)^{i*}$. If we define $(i*)$ as having area $\Delta A$, then it follows that:
\begin{equation}
 \frac{\partial}{\partial t} (bx)^{i*} = -\frac{1}{\Delta A} \int_A \nabla \times \vc{E} \cdot d\vc{A}.
\end{equation}
Note the use of the lower case $b$ to denote a face-averaged magnetic field quantity. (We will later use the upper-case $B$ to distinguish the \textit{cell}-averaged magnetic field quantities.) Then, using Stokes' theorem, we have: 
\begin{equation}
  \frac{\partial}{\partial t} (bx)^{i*} = - \frac{1}{\Delta A} \int_{\partial A} \vc{E} \cdot d\vc{r},
  \label{eqn:afterstokestheorem}
\end{equation}
where we now integrate over the path $\partial A$, which is the boundary that surrounds the face $(i*)$. We can discretise this equation, because we know this boundary comprises four edges in discretised space: $(i*, j*)$, $(i*, j*-1)$, $(i*, k*)$ and $(i*, k*-1)$. Furthermore, if we integrate in time, and use the fact that $\Delta A = \Delta y \Delta z$, then we obtain the following expression for $(bx)^{i*}_{n+1}$ given $(bx)^{i*}_{n}$:
\begin{equation}
	(bx)_{n+1}^{i*} = (bx)_{n}^{i*} - \displaystyle\frac{\Delta t}{\Delta y}\left[(Ez)^{i*, j*}_{n*} - (Ez)^{i*, j*-1}_{n*}\right] +\displaystyle \frac{\Delta t}{\Delta z}\left[(Ey)^{i*, k*}_{n*} - (Ey)^{i*, k*-1}_{n*}\right],
\label{eqn:facemagfield1}
\end{equation}
where $(Ez)^{i*, j*}$ is the $z$-component of the electric field, but averaged on the \textit{edge} $(i*, j*)$:
\begin{equation}
 (Ez)^{i*, j*} = \frac{1}{\Delta z} \int_{(i*, j*)} \vc{E}\cdot d\mathbf{z}.
 \label{eqn:edgeaverage}
\end{equation}
By substituting eq. (\ref{eqn:edgeaverage}) into eq. (\ref{eqn:facemagfield1}), it should be clear how the latter follows from eq. (\ref{eqn:afterstokestheorem}). Using similar reasoning, we can then derive expressions for the other two (face-averaged) components of the magnetic field, $(by)^{j*}_{n+1}$ and $(bz)^{k*}_{n+1}$: 
\begin{equation}
	(by)_{n+1}^{j*} = (by)_{n}^{j*}- \displaystyle \frac{\Delta t}{\Delta z}\left[(Ex)^{j*, k*}_{n*} - (Ex)^{j*, k*-1}_{n*}\right] + \displaystyle \frac{\Delta t}{\Delta x}\left[(Ez)^{i*, j*}_{n*}  - (Ez)^{i*-1, j*}_{n*}\right],
\label{eqn:facemagfield2}
\end{equation}
\begin{equation}
	(bz)_{n+1}^{k*} = (bz)_{n}^{k*}- \displaystyle \frac{\Delta t}{\Delta x}\left[(Ey)^{i*, k*}_{n*} - (Ey)^{i*-1, k*}_{n*}\right]+\displaystyle \frac{\Delta t}{\Delta y}\left[(Ex)^{j*, k*}_{n*} - (Ex)^{j*-1, k*}_{n*}\right].
\label{eqn:facemagfield3}
\end{equation}

\subsection{The electric field}
\label{sec:electricfield}
\begin{figure}
	\centering
	\includegraphics[width=\linewidth]{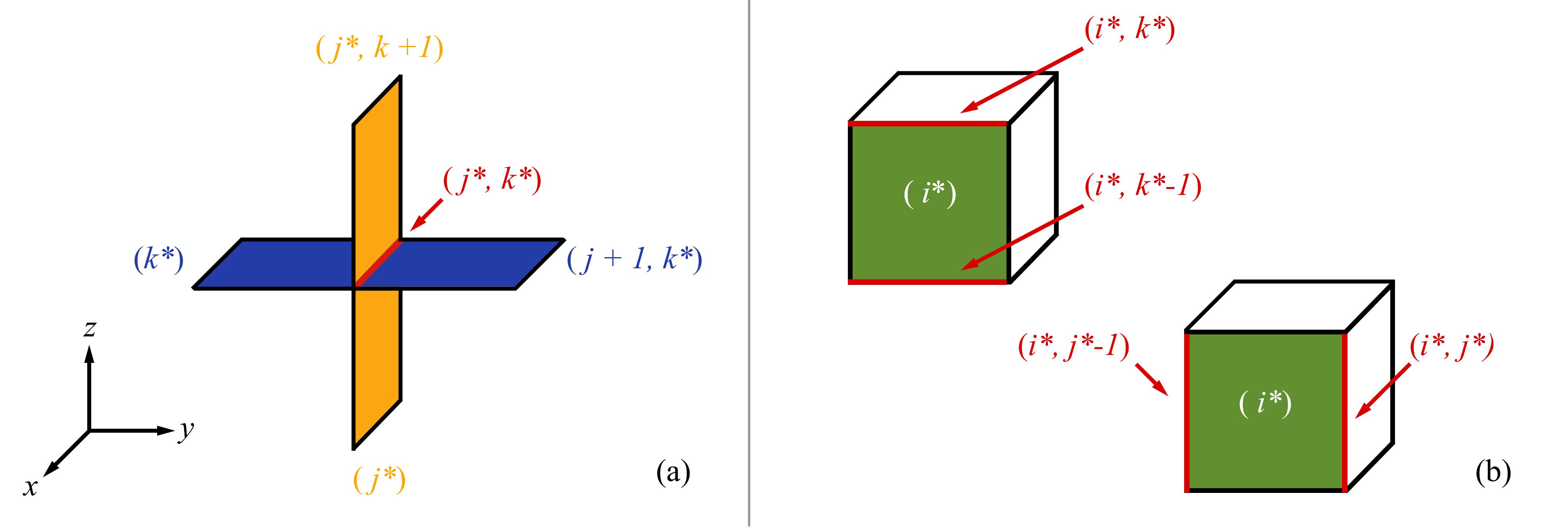}
	\caption{\textbf{(a) The edge-averaged electric field.} The edge $(j*, k*)$ is highlighted in red. To determine the edge average of $(Ex)$ along this edge, we take the arithmetic mean of the \textit{face} averages of $(Ex)$ on the surrounding four faces. These, in turn, are returned by the Riemann solver as inter-cell fluxes during the solve stage. See \S\ref{sec:electricfield} for more details. \textbf{(b) The face-averaged electric field.} We can then determine two different face-averaged components of the electric field on each face. For example, on the face $(i*)$ -- highlighted in green -- we can determine $(Ey)^{i*}_{\rightarrow x}$ by averaging $(Ey)$ along the edges $(i*, k*)$ and $(i*, k*-1)$ (top left), or $(Ez)^{i*}_{\rightarrow x}$ by averaging $(Ez)$ along the edges $(i*, j*)$ and $(i*, j*-1)$ (bottom right). See \S\ref{sec:cellaveragedmagneticfield} for more details.}
	\label{img:fourfaces}
\end{figure}

The next step is to determine the electric field components in eqs. (\ref{eqn:facemagfield1}), (\ref{eqn:facemagfield2})-(\ref{eqn:facemagfield3}). \citet[$\varphi = 1/\sqrt{4\pi}$]{BAL99} use the fact that these components appear in the last three rows of the MHD flux vectors -- given in eq. (\ref{eqn:fluxvectors}) -- as fluxes of the magnetic field quantities. This means that face-averaged values of these components will be returned by the Riemann solver during the solve stage. We can convert these into the \textit{edge}-averaged values we require by simple arithmetic averaging; to make the explanation of this clearer, we re-label the last three rows of the flux vectors as follows (where tildes signify that these are the values returned by the Riemann solver):
\begin{equation}
{\tilde\vc{F}}:\left[
	\begin{array}{c}
	 	0 \\
		(\tilde F7)_{n*}^{i*}\\
		(\tilde F8)_{n*}^{i*}
	\end{array}\right],
\quad {\tilde\vc{G}}:\left[
	\begin{array}{c}
	 	(\tilde G6)_{n*}^{j*} \\
		0 \\
		(\tilde G8)_{n*}^{j*}
	\end{array}\right],
\quad {\tilde\vc{H}}:\left[
	\begin{array}{c}
	 	(\tilde H6)_{n*}^{k*} \\
		(\tilde H7)_{n*}^{k*} \\
		0
	\end{array}\right].
\label{eqn:relabelledfluxes}
\end{equation}
So for example, $(Ex)$ appears as $(\tilde G8)_{n*}^{j*}$ on $y = $ constant faces and as $(\tilde H7)_{n*}^{k*}$ on $z = $ constant faces. In our grid of cells, we know that each edge is adjacent to four faces. For example, the edge $(j*, k*)$ is adjacent to the faces $(k*)$, $(j+1, k*)$, $(j*)$ and $(j*, k+1)$, as depicted in Fig. \ref{img:fourfaces}a. So to obtain the $x$-component of the electric field averaged along this edge, $(Ex)^{j*, k*}_{n*}$, Balsara and Spicer simply take the arithmetic mean of the appropriate fluxes through the surrounding four faces (where minus signs compensate for the minus signs in eq. (\ref{eqn:fluxvectors})):
\begin{equation}
  (Ex)^{ j*, k*}_{n*} = \frac{1}{4} \left[ (\tilde H7)_{n*}^{k*} + (\tilde H7)_{n*}^{ j+1, k*}- (\tilde G8)_{n*}^{j*} - (\tilde G8)_{n*}^{j*, k+1}\right],
 \label{eqn:edgeelectricfield1}
\end{equation}
and similarly for the other components:
\begin{equation}
  (Ey)^{ i*, k*}_{n*} = \frac{1}{4} \left[ (\tilde F8)_{n*}^{i*} + (\tilde F8)_{n*}^{i*, k+1}-(\tilde H6)_{n*}^{k*} - (\tilde H6)_{n*}^{i+1, k*}\right],
 \label{eqn:edgeelectricfield2}
\end{equation}
\begin{equation}
  (Ez)^{ j*, k*}_{n*} = \frac{1}{4} \left[ (\tilde G6)_{n*}^{j*} + (\tilde G6)_{n*}^{i+1, j*}- (\tilde F7)_{n*}^{i*} - (\tilde F7)_{n*}^{i*, j+1}\right],
 \label{eqn:edgeelectricfield3}
\end{equation}
This completes the description of the Balsara-Spicer algorithm.

\subsection{The cell-averaged magnetic field}
\label{sec:cellaveragedmagneticfield}
It is clear that the Balsara-Spicer algorithm requires a staggered grid: the magnetic field quantities are evolved on the cell faces, while the fluid quantities are evolved at the cell centres. Whenever a cell-averaged magnetic field is required (for example, during the reconstruction stage), then we must interpolate.  A linear interpolation is usually adequate:
\begin{equation}
  (Bx)_{n}^{i} = \frac{1}{2}\left[(bx)_{n}^{i*} + (bx)_{n}^{i*-1}\right],
\end{equation}
where we now use the upper-case $B$ to denote a cell-averaged magnetic field. \citet{TOT00} proposed a simple extension to this idea: to build the interpolation into the algorithm itself, thus eliminating the need for a staggered grid. His paper only includes the equations for a 2D system, but as he mentions, it is straightforward to extend these to three dimensions: 
\begin{equation}
 	(Bx)_{n+1}^{\bullet} = (Bx)_{n}^{\bullet}- \frac{\Delta t}{\Delta y}\left[( Ez)_{\rightarrow y}^{j*} - ( Ez)_{\rightarrow y}^{ j*-1}\right]+ \frac{\Delta t}{\Delta z}\left[( Ey)_{\rightarrow z}^{k*} - ( Ey)_{\rightarrow z}^{k*-1}\right],
\label{eqn:cellmagfield1}
\end{equation}
\begin{equation}
 	(By)_{n+1}^{\bullet} = (By)_{n}^{\bullet} - \frac{\Delta t}{\Delta z}\left[( Ex)_{\rightarrow z}^{k*} - ( Ex)_{\rightarrow z}^{k*-1}\right] + \frac{\Delta t}{\Delta x}\left[( Ez)_{\rightarrow x}^{i*} - ( Ez)_{\rightarrow x}^{i*-1}\right],
\label{eqn:cellmagfield2}
\end{equation}
\begin{equation}
 	(Bz)_{n+1}^{\bullet} = (Bz)_{n}^{\bullet} -  \frac{\Delta t}{\Delta x}\left[( Ey)_{\rightarrow x}^{i*} - ( Ey)_{\rightarrow x}^{i*-1}\right] + \frac{\Delta t}{\Delta y}\left[( Ex)_{\rightarrow y}^{ j*} - ( Ex)_{ \rightarrow y}^{j*-1}\right].
\label{eqn:cellmagfield3}
\end{equation}
which is similar to eqs. (\ref{eqn:facemagfield1}), (\ref{eqn:facemagfield2})-(\ref{eqn:facemagfield3}), except we now require \textit{face}-averaged electric field components.
Note that these are \textit{not} the same as the face-averaged components we obtained earlier from the Riemann solver. Instead, they are found by taking the arithmetic mean of the edge-averaged electric field components -- of eqs. (\ref{eqn:edgeelectricfield1})-(\ref{eqn:edgeelectricfield3}) -- in two surrounding, parallel edges. Each face has two such face-averaged electric field components, which depend on the two edges we use to calculate this mean. For example, for the face $(i*)$, we have: 
\begin{equation}
 (Ey)_{ \rightarrow x}^{i*} = \frac{1}{2}\left[(Ey)^{i*, k*} + (Ey)^{i*, k*-1}\right] \quad \rmn{and} \quad(Ez)_{ \rightarrow x}^{i*} = \frac{1}{2}\left[(Ez)^{i*, j*} + (Ez)^{i*, j*-1}\right].
\end{equation}
Both of these are illustrated in Fig. \ref{img:fourfaces}b. We use the $\rightarrow$ subscript to distinguish, for example, between $(Ey)_{ \rightarrow x}^{i*}$ through the $x = $ constant faces and and $(Ey)_{ \rightarrow z}^{k*}$ through the $z = $ constant faces. Using eqs. (\ref{eqn:edgeelectricfield1})-(\ref{eqn:edgeelectricfield3}), we then have:
\begin{equation}
     ( Ey)_{ \rightarrow x}^{i*} =  \frac{1}{8} \left[(\tilde F8)_{n*}^{i*, k} + (\tilde F8)_{n*}^{i*, k+1} - (\tilde H6)_{n*}^{i,  k*} - (\tilde H6)_{n*}^{i+1, k*} + (\tilde F8)_{n*}^{i*, k-1} + (\tilde F8)_{n*}^{i*, k} - (\tilde H6)_{n*}^{i, k*-1} - (\tilde H6)_{n*}^{i+1, k*-1}\right],
     \label{eqn:eyx}
\end{equation}
\begin{equation}
    ( Ez)_{ \rightarrow x}^{i*} =   \frac{1}{8} \left[(\tilde G6)_{n*}^{i, j*} + (\tilde G6)_{n*}^{i+1, j*} - (\tilde F7)_{n*}^{i*, j} - (\tilde F7)_{n*}^{i*, j+1} + (\tilde G6)_{n*}^{i, j*-1} + (\tilde G6)_{n*}^{i+1, j*-1} - (\tilde F7)_{n*}^{i*, j-1} - (\tilde F7)_{n*}^{i*, j}\right],
    \label{eqn:ezx}
\end{equation}
In a similar manner, we can derive the face-averaged components through the faces $(i*)$ and $(k*)$: 
\begin{equation}
     ( Ex)_{ \rightarrow y}^{j*} =\frac{1}{8} \left[(\tilde H7)_{n*}^{ j, k*} + (\tilde H7)_{n*}^{j+1, k*} - (\tilde G8)_{n*}^{j*, k} - 	(\tilde G8)_{n*}^{j*, k+1}+ (\tilde H7)_{n*}^{j, k*-1} + (\tilde H7)_{n*}^{j+1, k*-1} - (\tilde G8)_{n*}^{ j*, k-1} - 	(\tilde G8)_{n*}^{j*, k}\right],
 \label{eqn:exy}
\end{equation}
\begin{equation}
      ( Ez)_{ \rightarrow y}^{j*} =  \frac{1}{8} \left[(\tilde G6)_{n*}^{i, j*} + (\tilde G6)_{n*}^{i+1, j*} - (\tilde F7)_{n*}^{i*, j} - (\tilde F7)_{n*}^{i*, j+1} +(\tilde G6)_{n*}^{i-1, j*} + (\tilde G6)_{n*}^{i, j*} - (\tilde F7)_{n*}^{i*-1, j} - (\tilde F7)_{n*}^{i*-1, j+1}\right].
    \label{eqn:ezy}
\end{equation}
\begin{equation}
    ( Ex)_{ \rightarrow z}^{k*} =  \frac{1}{8} \left[(\tilde H7)_{n*}^{ j, k*} + (\tilde H7)_{n*}^{j+1, k*} - (\tilde G8)_{n*}^{j*, k} - 	(\tilde G8)_{n*}^{j*, k+1} + (\tilde H7)_{n*}^{ j-1, k*} + (\tilde H7)_{n*}^{j, k*} - (\tilde G8)_{n*}^{j*-1, k} - 	(\tilde G8)_{n*}^{j*-1, k+1}\right],
   \label{eqn:exz}
\end{equation}
\begin{equation}
  ( Ey)_{ \rightarrow z}^{k*} =  \frac{1}{8} \left[(\tilde F8)_{n*}^{i*, k} + (\tilde F8)_{n*}^{i*, k+1} - (\tilde H6)_{n*}^{i, k*} - (\tilde H6)_{n*}^{i+1, k*} + (\tilde F8)_{n*}^{i*-1, k} + (\tilde F8)_{n*}^{i*-1, k+1} - (\tilde H6)_{n*}^{i-1, k*} - (\tilde H6)_{n*}^{i, k*}\right].
   \label{eqn:eyz}
\end{equation}
This concludes the description of the Balsara-Spicer-T\'oth algorithm.

\subsection{Disadvantages of this algorithm}
The simplest implementation of this algorithm involves creating six new 3D arrays, one for each of the inter-cell fluxes we require: $(\tilde F7)$, $(\tilde F8)$, $(\tilde G6)$, $(\tilde G8)$, $(\tilde H6)$ and $(\tilde H7)$. Every time we complete the solve stage for a given 1D row, we obtain the values of two of these fluxes along the entire row, and so can store them in the appropriate arrays. (The two we obtain depend on what the directional pass is.) Once a complete set of three directional passes has been completed, we can sort these arrays into another six 3D arrays, one for each of the face-averaged electric field components: $(Ex)_{ \rightarrow y}^{j*}$, $(Ex)_{ \rightarrow z}^{k*}$, $(Ey)_{ \rightarrow x}^{i*}$, $(Ey)_{ \rightarrow z}^{k*}$, $(Ez)_{ \rightarrow x}^{i*}$ and $(Ez)_{ \rightarrow y}^{j*}$. This is done according to eqs. (\ref{eqn:eyx})-(\ref{eqn:eyz}) above. Finally, we use these components to update the magnetic field quantities using eqs. (\ref{eqn:cellmagfield1})-(\ref{eqn:cellmagfield3}). 

This algorithm has been shown to be reliable through standard MHD tests \citep{TOT00}, but has some notable disadvantages: 
\begin{enumerate}
\item \textit{Storage of information.} In a DS-RSA hydrodynamics code, we can extract a single row from the 3D grid, update all of the quantities along that row, and then return it. We do not need to store any information from that row to evolve other rows in the same directional pass. This is clearly not the case here: we need to store all of the inter-cell fluxes and electric field components.
\item \textit{Processing between sets of directional passes.} In a DS-RSA hydrodynamics code, once a complete set of directional passes has been completed, the state variables require no further operations until the next set. Again, this is not the case here: we need to carry out additional operations on the inter-cell fluxes and electric field components to determine the new magnetic field quantities. 
\item \textit{Staggered grids.} Although the magnetic field is stored as cell averages, we still require staggered grids: the inter-cell fluxes and electric field components are stored as \textit{face} averages. This complicates their implementation, especially with adaptive/non-uniform grids.
\item \textit{Limited parallelisation.} Finally, because of the extra processing between directional passes, and the fact that the electric field components depend on inter-cell fluxes from different rows \textit{and} directional passes, it is not immediately obvious how this code could be parallelised.
\end{enumerate}

While these are arguably not major issues, they complicate the implementation of this algorithm in existing, mature astrophysical codes. We would therefore prefer an algorithm that: minimises the amount of data that needs to be stored from each row during the directional passes; reduces the amount of processing required between complete sets of directional passes; completely removes the need for any staggered grids; simplifies and maximises the possible parallelisation. We now show that such an algorithm exists, by reformulating the Balsara-Spicer-T\'oth algorithm so that it is more compatible with a dimensionally-split framework.

\section{A Modified Approach}

To start, we define just three new 3D arrays: $\Delta (Bx)$, $\Delta (By)$ and $\Delta (Bz)$. At the end of each complete set of three directional passes, we aim for these to contain the change in the magnetic field during those passes. To make our derivation clearer, we focus on a single cell: $(i, j, k)$.

\subsection{The $x$-direction pass}

\subsubsection{Contribution to $\Delta (Bx)^\bullet$}
The change in $\Delta (Bx)^\bullet$ during each timestep is given by eq. (\ref{eqn:cellmagfield1}), and depends on the components $(\bar Ez)^{j*}_{ \rightarrow y}$, $(\bar Ez)^{j*-1}_{ \rightarrow y}$, $(\bar Ey)^{k*}_{ \rightarrow z}$ and  $(\bar Ey)^{k*-1}_{ \rightarrow z}$. We can substitute in the definition of these components, given by eqs. (\ref{eqn:eyz}) and (\ref{eqn:ezy}). However, we only include the $\tilde{\vc{F}}$ terms, as these are the only ones we obtain during the $x$-direction pass. Then, after simplification, we are left with the following expression:
\begin{equation}
  \begin{array}{ll}
      \Delta (Bx)^{\bullet} := \Delta (Bx)^{\bullet} &+ \frac{\Delta t}{\Delta y} \frac{1}{8}\left[(\tilde F7)_{n*}^{i*, j+1} + (\tilde F7)_{n*}^{i*-1, j+1} - (\tilde F7)_{n*}^{i*, j-1} - (\tilde F7)_{n*}^{i*-1, j-1} \right]\\ 
      &+ \frac{\Delta t}{\Delta z}\frac{1}{8} \left[(\tilde F8)_{n*}^{i*, k+1} + (\tilde F8)_{n*}^{i*-1, k+1} - (\tilde F8)_{n*}^{i*, k-1} - (\tilde F8)_{n*}^{i*-1, k-1}\right],
  \end{array}
 \label{eqn:deltabxupdate}
\end{equation} 
where use `$:=$' to refer to an update in the usual computational sense.

This is still not in the spirit of dimensionally-split algorithms, as the right-hand side of the expression depends on fluxes from several different rows, as shown by the different spatial indices. Instead, we want to consider what we could do after a \textit{given} $x$-row -- labelled $(i, J, K)$ with fixed $J, K$ -- has been evolved. To do this, we use the fact that if $\Delta (Bx)^\bullet$ depends on, say, $(\tilde F7)_{n*}^{i*, j+1}$, then this is equivalent to saying that  $\Delta (Bx)^{j-1}$ depends on $(\tilde F7)_{n*}^{i*}$. If we consider each term in eq. (\ref{eqn:deltabxupdate}) in this manner, then we can rewrite it as: 
\begin{equation}
  \Delta (Bx)^{J\pm1} := \Delta (Bx)^{J\pm1} \mp \frac{\Delta t}{\Delta y} \frac{1}{8} \left[(\tilde F7)_{n*}^{i*} + (\tilde F7)_{n*}^{i*-1}\right],
  \label{eqn:newmethod-bx1}
\end{equation}
\begin{equation}
  \Delta (Bx)^{K\pm1} :=  \Delta (Bx)^{K\pm1} \mp \frac{\Delta t}{\Delta z} \frac{1}{8} \left[(\tilde F8)_{n*}^{i*} + (\tilde F8)_{n*}^{i*-1}\right].
  \label{eqn:newmethod-bx2}
\end{equation}
Both the right-hand sides only contain terms from the row we are currently on, $(i*)$. This means that once a given row has been evolved, we can process the contribution of the inter-cell fluxes to the magnetic field quantities immediately -- using eqs. (\ref{eqn:newmethod-bx1}) and (\ref{eqn:newmethod-bx2}) -- without needing to store them in any other form. It is also worth noting that these equations are near identical to the standard RSA approach given in eq. (\ref{eqn:rsasteps}). 

\subsubsection{Contribution to $\Delta (By)^\bullet$ and $\Delta (Bz)^\bullet$}
Following a similar approach, we use eq. (\ref{eqn:cellmagfield2}) to find the contribution to $\Delta (By)^\bullet$ during the $x$-direction pass:
\begin{equation}
    \Delta (By)^{\bullet} := \Delta (By)^{\bullet} + \frac{\Delta t}{\Delta x} \frac{1}{4} \left[- (\tilde F7)_{n*}^{i*} + (\tilde F7)_{n*}^{i*-1}\right],
\end{equation}
\begin{equation}
    \Delta (By)^{j\pm1} := \Delta (By)^{j\pm1} + \frac{\Delta t}{\Delta x} \frac{1}{8} \left[- (\tilde F7)_{n*}^{i*} + (\tilde F7)_{n*}^{i*-1}\right],
\end{equation}
while eq. (\ref{eqn:cellmagfield3}) gives us the contribution to $\Delta (Bz)^\bullet$ during the $x$-direction pass:
\begin{equation}
    \Delta (Bz)^{\bullet} := \Delta (Bz)^{\bullet} + \frac{\Delta t}{\Delta x} \frac{1}{4} \left[- (\tilde F8)_{n*}^{i*} + (\tilde F8)_{n*}^{i*-1}\right],
\end{equation}
\begin{equation}
    \Delta (Bz)^{k\pm1} := \Delta (Bz)^{k\pm1} + \frac{\Delta t}{\Delta x} \frac{1}{8} \left[- (\tilde F8)_{n*}^{i*} + (\tilde F8)_{n*}^{i*-1}\right].
    \label{eqn:newmethod-bz2}
\end{equation}

\subsection{The $y$- and $z$-direction passes}
We go through the same steps as above for the $y$-direction pass, but start by considering $\Delta (By)^\bullet$. The change in this during each timestep is given by eq. (\ref{eqn:cellmagfield2}), and depends on the components $(\bar Ex)^{k*}_{ \rightarrow z}$, $(\bar Ex)^{k*-1}_{ \rightarrow z}$, $(\bar Ez)^{i*}_{ \rightarrow x}$ and  $(\bar Ez)^{i*-1}_{ \rightarrow x}$. Again, we substitute in the definition of these components, given by eqs. (\ref{eqn:exz}) and (\ref{eqn:ezx}), although this time we only include the $\tilde\vc G$ terms. After simplification, we are left with: 
\begin{equation}
   \begin{array}{ll}
      \Delta (By)^{\bullet} :=  \Delta (By)^{\bullet} &+ \frac{\Delta t}{\Delta x} \left[(\tilde G6)_{n*}^{i+1, j*} + (\tilde G6)_{n*}^{i+1, j*-1} - (\tilde G6)_{n*}^{i-1, j*} - (\tilde G6)_{n*}^{i-1, j*-1} \right] \\
      & + \frac{\Delta t}{\Delta z} \left[(\tilde G8)_{n*}^{j*, k+1} + (\tilde G8)_{n*}^{j*-1, k+1} - (\tilde G8)_{n*}^{j*, k-1} - (\tilde G8)_{n*}^{j*-1, k-1}\right].
 \end{array}
 \label{eqn:deltabyupdate}
\end{equation} 
Before we continue, a note on dimensionally-split codes: when we extract a given row from each full 3D array (of which there is one for each state quantity), we store them in corresponding 1D arrays. However, the quantities that are broken down into Cartesian components - the velocity and the magnetic field - must be extracted differently depending on the directional pass. For example, in the $x$-direction pass, $(vx)$ is parallel to the row, and is stored in a 1D array we shall call $(vx)^{\rmn{1D}}$. However, in the $y$-direction pass, it is $(vy)$ that is parallel to the row. To avoid having to rewrite the 1D algorithms for each directional pass, we simply store $(vy)$ in the array $(vx)^{\rmn{1D}}$. This approach is summarised in Table \ref{tbl:dimsplit}. If we rewrite the terms in eq. (\ref{eqn:deltabyupdate}) according to this approach -- such that $\{x, F, i, 6\} \rightarrow \{y, G, j, 7\}$ and $\{y, G, j, 7\} \rightarrow \{x, F, i, 6\}$ -- then we actually retrieve eq. (\ref{eqn:deltabxupdate}). Furthermore, it can be shown that \textit{all} of eqs. (\ref{eqn:newmethod-bx1})-(\ref{eqn:newmethod-bz2}) can be retrieved in this manner, and for the $z$-direction pass as well. In other words, once the equations for the $x$-direction pass has been implemented, we do not need to re-write them for the $y$- and $z$-direction passes, as long as the Cartesian components are extracted and returned as described above. 

\begin{table}
 	 \caption{In a dimensionally-split code, we can avoid rewriting the 1D algorithms if we make sure that the Cartesian components of the velocity are extracted (and then returned) according to this table. A similar strategy applies for the components of the magnetic field.}
 	 \label{tbl:dimsplit}
 	 \begin{tabular}{lc c c }
	\hline
	Dimensional pass & $\mathbf{x}$ &$\mathbf{y}$ & $\mathbf{z}$\\
	\hline
	$(vx)^{\rmn{1D}}$ stores & $(vx)$ & $(vy)$ & $(vz)$ \\
	$(vy)^{\rmn{1D}}$ stores & $(vy)$ & $(vx)$ & $(vx)$\\
	$(vz)^{\rmn{1D}}$ stores & $(vz)$ & $(vz)$ & $(vy)$ \\
	\hline
 	 \end{tabular}
\end{table}

\subsection{Final update}
Once all three passes have been completed, we simply need to perform the following operation:
 \begin{equation}
  (Bx) := (Bx) + \Delta (Bx)
  \label{eqn:magfieldupdate}
 \end{equation}
 and similarly for $(By)$ and $(Bz)$ to update the magnetic field. There are some additional subleties associated with this, regarding the pressure and energy (which depend on the magnetic field), and these are described in the next section.

 \subsection{Comparison with the Balsara-Spicer-T\'oth Algorithm}
 \begin{enumerate}
  \item \textit{Reduced storage of information.} We have minimised the amount of information that needs to be stored during the directional passes. Instead of storing the individual inter-cell fluxes and electric field components (which requires twelve 3D arrays), we only store the change in the magnetic field quantities (which requires just three). 
  \item \textit{Reduced processing between sets of directional passes.} We have also minimised the number of operations required between complete sets of directional passes: we only need to perform the simple calculation given in eq. (\ref{eqn:magfieldupdate}).
  \item \textit{No staggered grids.} Furthermore, the three arrays we require store only \textit{cell}-averaged values, which means we no longer need to store any information on the cell faces. This simplifies the use of adaptive and non-uniform grids.
  \item \textit{Simplified parallelisation.} The modified algorithm is more easily parallelisable. For example, assume that we have a number of processors, each independently evolving some rows in a given directional pass. Each of these processors should have a copy of the $\Delta \vc B$ array, which are updated as described above. Then, because these all contain \textit{changes} in the magnetic field quantities, we only require a simple `collect'-style operation to sum over all of these copies and obtain the total change before completing eq. (\ref{eqn:magfieldupdate}).
  \item \textit{Dimensionally-split framework.} Finally, because this algorithm is more compatible with the dimensionally-split framework, it is considerably simpler to include in existing dimensionally-split hydrodynamics codes.
 \end{enumerate}

\section{The Complete Algorithm}
\label{sec:algorithm}
We will now describe the complete algorithm for each timestep. Note that we require the following sets of arrays: 
\begin{enumerate}
 \renewcommand{\theenumi}{(\arabic{enumi})}
 \item $7 \times \vc{U}^{\rmn{3D}}$: the 3D arrays which store the cell-averaged state quantities as listed in eq. (\ref{eqn:statevector}). 
 \item $7 \times \vc{U}^{\rmn{1D}}$: the 1D arrays which are used to store 1D rows extracted from the full 3D grid. 
\end{enumerate}
Additionally, we require two more sets of arrays for the magnetic field quantities:
\begin{enumerate}
 \renewcommand{\theenumi}{(\arabic{enumi})}
 \setcounter{enumi}{2}
 \item $3 \times \vc{B}^{\rmn{s}}$: used to store a copy of the cell-averaged magnetic field quantities at the start of each timestep.
 \item $3 \times \Delta \vc{B}$: used to store the change in the cell-averaged magnetic field quantities during each timestep.
\end{enumerate}

\subsection{Before the directional passes}
\begin{enumerate} 
  \alphen{0}
  \item \textit{Make a copy of the magnetic field quantities.} Copy the three $\vc{U}^{\rmn{3D}}$ arrays that store the cell-averaged magnetic field quantities into the corresponding $\vc{B}^{\rmn{s}}$ arrays. These copies should not be modified during the directional passes. 
\end{enumerate}

\subsection{The $x$-direction pass}
\begin{enumerate}
  \alphen{1}
  \item \textit{Extract a 1D row.} Choose an $x$-direction row from the grid, which is defined as having fixed $j$ and $k$ indices. Extract the cell-averaged state quantities along this row from the $\vc{U}^{\rmn{3D}}$ arrays, and store them in the corresponding $\vc{U}^{\rmn{1D}}$ arrays. 
  \item \textit{Reconstruct the cell-averaged data.} For each $\vc{U}^{\rmn{1D}}$ array, interpolate the cell-averaged data to obtain left and right states at each cell interface $(i*)$. These will be input states for Riemann problems. For the tests in this paper, we use the TVD scheme of \citet[$\varphi = 1\sqrt{4\pi}$]{BAL98}, which includes a steepening algorithm. However, a popular alternative is the Piecewise Parabolic Method of \citet{COL84}, an MHD version of which is summarised in \S4.2.3 of \citet[$\varphi = 1$]{STO08}. 
  \item {Find the inter-cell fluxes.} Solve the Riemann problem at each cell interface, to obtain the inter-cell flux for each state variable. We use the multi-state HLLD approximate Riemann solver of \citet[$\varphi = 1$]{MIY05}.\footnote{A note on using non-staggered grids: in a 1D system, the parallel component of the magnetic field -- which is $(Bx)$ in the $x$-direction pass -- should be uniform and constant in time. This automatically satisfies the divergence-free condition. The reconstruction and solver algorithms we use were designed for such 1D systems, which means that the former does not find left and right states for this parallel component in step (c), while the latter assumes it is constant across the cell interface in step (d). This is not an issue when using staggered grids, because this constant value is stored explicitly at the cell interface. However, this is clearly not the case for non-staggered grids, which means we must interpolate this parallel component to get its value across the interface. Our tests have shown that a simple linear interpolation is sufficient, for example: $(Bx)_{\rmn{int}, i*} = [(Bx)_i + (Bx)_{i+1}]/2$.} 
  \item \textit{Update the state variables.} Using the inter-cell fluxes $\tilde\vc F$ which were determined in step (d), update the cell averages stored in $\vc{U}^{\rmn{1D}}$  using eq. (\ref{eqn:rsasteps}). Note that this should be done for the magnetic field quantities as well.\footnote{\label{ftn:gaspressure} This is because we require the \textit{gas} pressure $p_{\rmn{G}}$ when we return to the earlier steps. However, recall that we only store the total pressure $p_{\rmn{T}}$, so we need to calculate it by subtracting the magnetic pressure $p_{\rmn{M}}$ (i.e.$p_{\rmn{G}} = p_{\rmn{T}} - p_{\rmn{M}}$), which is in turn determined by the magnetic field quantities. Hence, in cases where $p_{\rmn{M}} \gg p_{\rmn{G}}$, not updating the magnetic field components can result in negative pressures. This issue is discussed further in \citet{BAL99}.}
\item \textit{Return the 1D row.} Return the new cell averages stored in the $\vc{U}^{\rmn{1D}}$ arrays to the $\vc{U}^{\rmn{3D}}$ arrays (again, including magnetic field quantities).
\item \textit{Complete the $x$-direction pass:} repeat steps (b) to (f) for each 1D row in the $x$-direction (that is, for all fixed $j, k$).
\end{enumerate}

\subsection{The $y$- and $z$-directional passes}
\begin{enumerate}
  \alphen{7}
  \item \textit{Complete all passes:} repeat the equivalents of steps (b) to (g) for the $y$- and $z$- directions. Remember to extract and return the velocity and magnetic field components in the usual dimensionally-split manner, as described in Table \ref{tbl:dimsplit}. Also note that in some code implementations, the order of the passes is alternated between timesteps to preserve the accuracy of the 1D method.
\end{enumerate}

\subsection{After the dimensional passes}
After all of the above steps have been completed, the fluid quantities will be fully evolved to the next timestep. We will also have three versions of the magnetic field quantities: their values at the start of the timestep (stored in $\vc{B}^{\rmn{s}}$), their (correct) change during the timestep as determined by the constrained transport algorithm (stored in $\Delta \vc{B}$), and their values as evolved -- incorrectly -- through the standard RSA strategy (stored in the $\vc{U}^{\rmn{3D}}$ arrays corresponding to the magnetic field quantities).  Then the final steps are: 

\begin{enumerate}
  \alphen{8}
  \item \textit{Correct the energy and pressure (I).} This is an optional step. We use the magnetic field quantities stored in the $\vc{U}^{\rmn{3D}}$ arrays to calculate the magnetic field pressure $p_{\rmn{M}}$ (using $p_{\rmn{M}} = \frac{1}{2} \varphi^2 \vc{B}^2$). We then subtract this contribution from the newly evolved values of the total energy (using the equation of state defined in eq. (\ref{eqn:equationofstate}) as a guide) and the total pressure. 
  \item \textit{Update the magnetic field}. We replace the magnetic field quantities stored in the $\vc{U}^{\rmn{3D}}$ arrays with $\vc{B}^{\rmn{s}}$ + $\Delta \vc{B}$ (in other words, their value at the start of the timestep, plus their \textit{correct} change in value during the timestep, according to the constrained transport algorithm). Consequently, the stored magnetic field quantities now maintain the divergence-free condition.
  \item \textit{Correct the energy and pressure (II).} If step (i) was completed: use these new magnetic field quantities to recalculate the magnetic pressure and add this back to the total energy and total pressure. This correction means that energy may not be conserved entirely \citep[as discussed in][]{BAL99}, but avoids negative pressures if $p_{\rmn{M}} \gg p_{\rmn{G}}$, as described in footnote \ref{ftn:gaspressure}. 
\end{enumerate}

\section{MHD Tests}

To illustrate the accuracy of the modified algorithm, we now present the results of four standard MHD tests. All results were obtained using the second-order TVD reconstruction of \cite{BAL98}, and the HLLD Riemann solver of \cite{MIY05}. Also note that we use units of $\varphi = 1/\sqrt{4\pi}$, except for the circularly-polarised Alfv\'en waves, which use $\varphi = 1$. 

\subsection{Circularly-polarised Alfv\'en waves}
\label{sec:alfven}
The test of circularly polarised Alfv\'en waves were first described by \citet{TOT00}. Such waves are smooth, analytic, non-linear solutions of the MHD equations, which makes them ideal for testing the implementation of these equations. The version implemented is sinusoidal and uses a periodic domain that can fit exactly one wavelength. We test two versions: a travelling wave, where the wave returns to its original position at integer multiples of the period (which is equal to 1 in this case), and a standing wave, where it stays at its original position. 

We use the initial conditions described in \citet{TOT00}. We start by defining a rotated coordinate system, $c_\parallel$ and $c_\bot$. These are rotated by an angle $\theta = \pi/6$ from the Cartesian $x$ and $y$ (and hence our discretised grid) such that: 
\begin{equation}
  c_\parallel = x \cos \theta + y \sin \theta \quad \rmn{and} \quad c_\bot = -x \sin \theta + y \cos \theta.
  \label{eqn:coordinatechange}
\end{equation}
We then define the velocity and magnetic field components along these axes: $v_\parallel = 0$ for travelling waves and $v_\parallel =1$ for standing waves; $B_\parallel = 1$; $v_\bot = B_\bot = 0.1 \sin (2\pi c_\parallel)$. We then have:
\begin{enumerate}
  \alphen{0}
  \item \textit{Space and time.} Domain: $0 < x < 1/\cos \theta$, $0 < y < \sin \theta$; number of cells: varying from $8^2$ to $32^2$; total time $t = 5$.
  \item \textit{Fluid quantities.} Across the entire grid: mass density $\rho = 1$; gas pressure $p_G = 0.1$; $x$-velocity $v_x = v_\parallel \cos \theta - v_\bot \sin \theta$; $y$-velocity $v_y = v_\parallel \sin \theta + v_\bot \cos \theta$; $z$-velocity $v_z = 0.1 \cos (2\pi c_\parallel)$; adiabatic index $\gamma = 5/3$. 
  \item \textit{Magnetic field quantities.} The magnetic field is set up equivalently to the velocity. Across the entire grid: $\varphi B_x = B_\parallel \cos \theta - B_\bot \sin \theta$; $\varphi B_y = B_\parallel \sin \theta + B_\bot \cos \theta$; $\varphi B_z = 0.1 \cos (2\pi c_\parallel)$. 
\end{enumerate}

 \begin{figure}
   \centering
   \subfigure[Results for travelling waves, $v_\parallel = 0$.]{
     \includegraphics[width=0.45\linewidth]{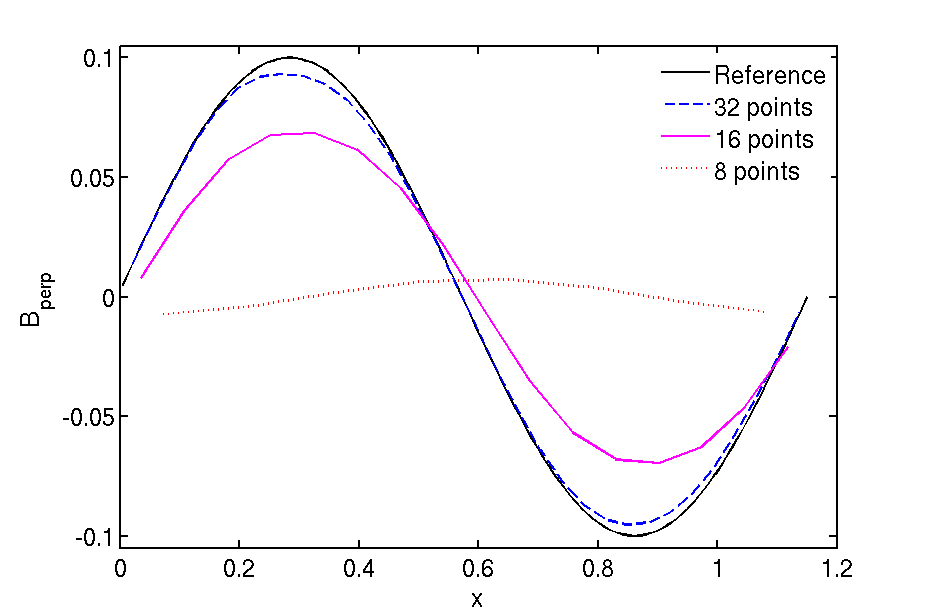}}
   \subfigure[Results for standing waves, $v_\parallel = 1$.]{
     \includegraphics[width=0.45\linewidth]{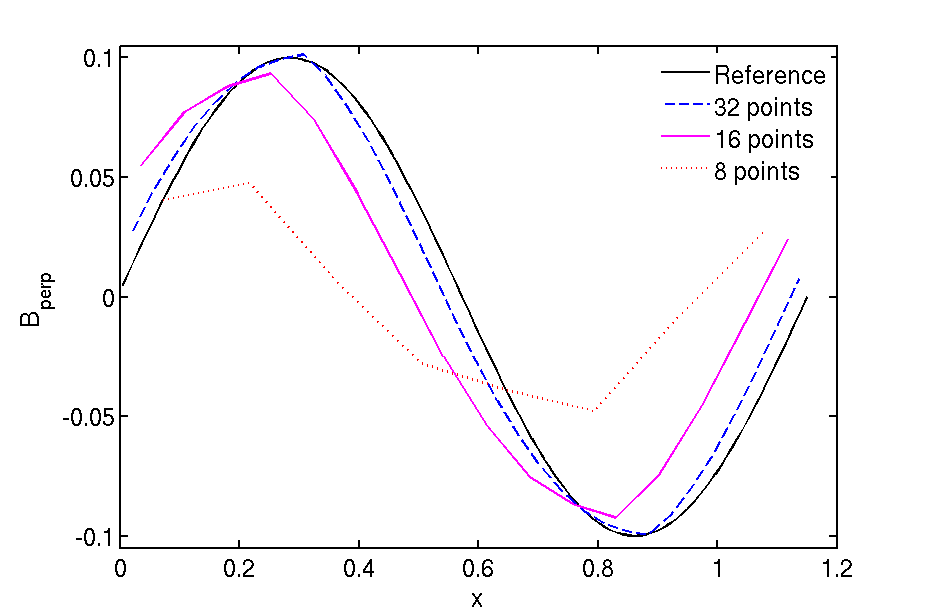}}
   \caption{Plot of $B_\perp = (\sqrt{3}B_y - B_x)/2$, from the circularly-polarised Alfv\'en waves test at time $t = 5$, initialised as described in \S\ref{sec:alfven}. }
   \label{fig:alfven-results}
 \end{figure}

The results for both the travelling and standing waves at time $t = 5$ are shown in Fig. \ref{fig:alfven-results}, and can be compared with Figs. 8 and 9 in \citet{TOT00}. Note that we plot $B_\perp$, which can be determined from $B_x$ and $B_y$ using eq. (\ref{eqn:coordinatechange}). We include reference plots, which are simply the initial conditions on a grid with 128$\times$128 cells. This was designed as a difficult test -- especially at low resolutions -- so the results do not coincide exactly with the expected analytic solution. However, they are comparable with results from other algorithms, as presented in \citet{TOT00}. It may also be possible to improve them with the use of a higher-order reconstruction.

\subsection{The blast problem}
\label{sec:blast}
The blast problem is initialised as a small disk of high-pressure fluid in the centre of the grid, which then expands rapidly into the low-pressure ambient fluid. As pointed out by \citet{BAL99}, it is not necessarily useful for testing whether the magnetic field is divergence free: the conditions are such that the build-up of non-zero $\nabla \cdot \vc B$ will not have a major effect on the overall dynamics. However, it is useful for testing the propagation of strong MHD shocks in multiple dimensions \citep{STO08}. 


 We use the initial conditions described in \citet{STO08} \citep[which were in turn taken from][]{LON00}: 
 \begin{enumerate}
 \alphen{0}
  \item \textit{Space and time.} Domain: $-0.5 < x < 0.5$; $-0.75 < y < 0.75$; number of cells: 300 $\times$ 450; total time $t = 0.2$.
  \item \textit{Fluid quantities.} Across the entire grid: mass density $\rho = 1$; velocity $\vc v = 0$; adiabatic index $\gamma = 5/3$. The gas pressure is set to $p_G = 10$ within a radius $r < 0.1$ in the centre of the grid, and $p_G = 0.1$ everywhere else. 
  \item \textit{Magnetic field quantities.} Across the entire grid: $\varphi B_x = \varphi B_y = 1/\sqrt{2}$; $\varphi B_z = 0$. 
 \end{enumerate}

 \begin{figure}
   \centering
   \subfigure[Mass density.]{
     \includegraphics[width=0.3\linewidth]{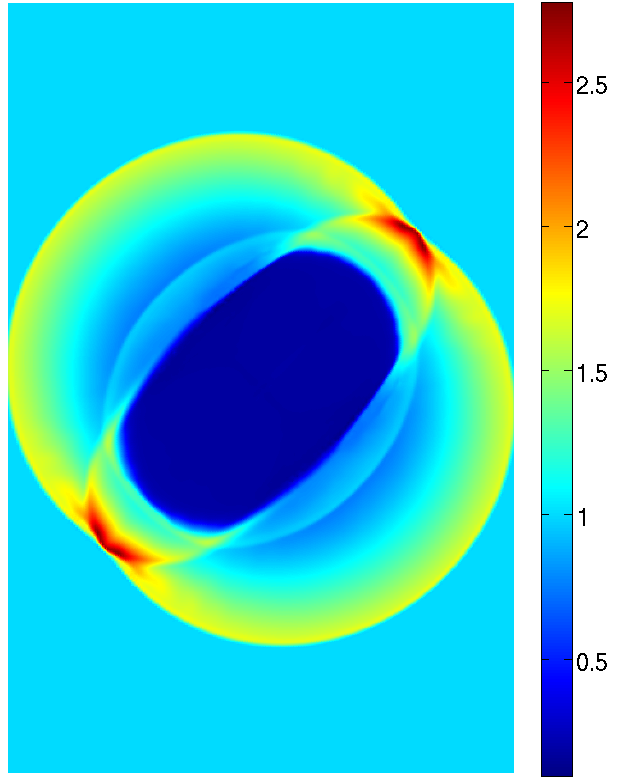}}
   \subfigure[Thermal pressure.]{
     \includegraphics[width=0.3\linewidth]{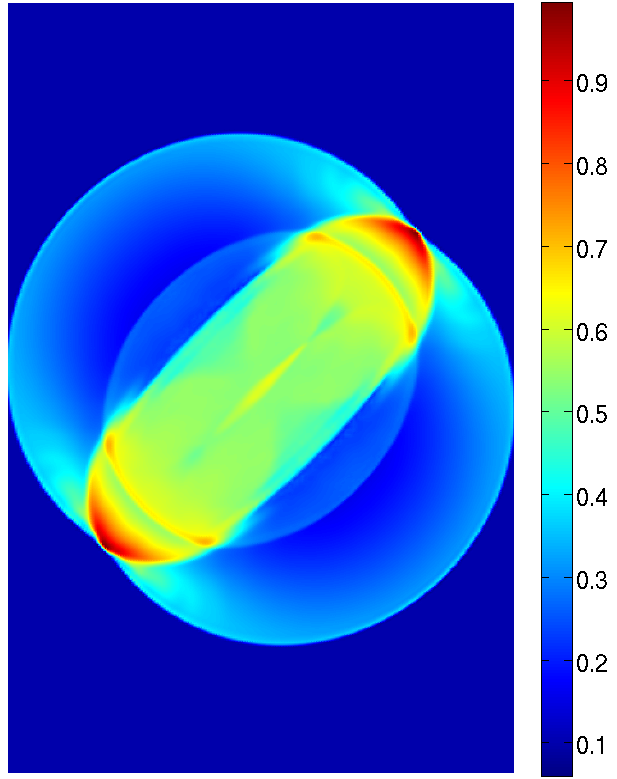}}
   \subfigure[Magnetic pressure.]{
     \includegraphics[width=0.3\linewidth]{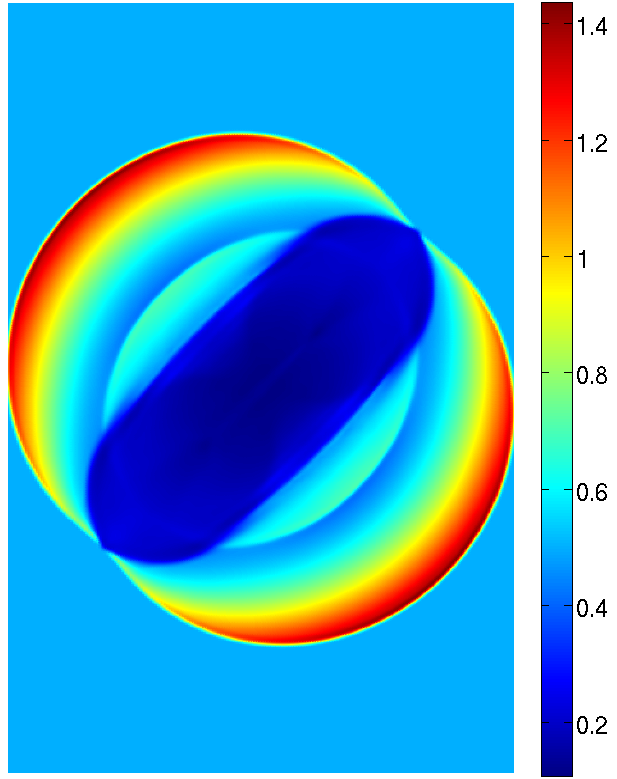}}
   \caption{Results from the blast problem at time $t = 0.2$, initialised as described in \S\ref{sec:blast}. }
   \label{fig:blast-results}
 \end{figure}

The mass density, thermal pressure and magnetic pressure at time $t = 0.2$ are shown in Fig. \ref{fig:blast-results}, and can be compared with Fig. 28 in \citet{STO08}. The results are as expected: the blast -- which would otherwise be circular - has aligned itself with the magnetic field. The near-perfect $\pi$-rotational symmetry of the results are also of note.

\subsection{The Orszag-Tang vortex}
\label{sec:orszag}
The Orszag-Tang vortex is another commonly-used 2D MHD test, which starts with smooth initial data but becomes progressively more complex \citep{TOT00}. The expected symmetry of the solution is useful for showing that the dimensionally-split algorithm (which uses the same code for each dimensional pass, but just extracts and returns the rows differently) works and has been implemented correctly. 

 We use the initial conditions described in \citet{STO08} \citep[which were in turn taken from][]{ORS79}.
 \begin{enumerate}
  \alphen{0}
  \item \textit{Space and time.} Domain: $0 < x, y < 1$; number of cells: $128^2$; total time $t = 0.5$. 
  \item \textit{Fluid quantities.} Across the entire grid: mass density $\rho = 25/36\pi$; pressure $p = 5/12\pi$;  $x$-velocity $v_x = -\sin(2\pi y)$; $y$-velocity $v_y = \sin(2\pi x)$; $z$-velocity $v_z = 0$; adiabatic index $\gamma = 5/3$. 
  \item \textit{Magnetic field quantities.} Across the entire grid: $\varphi B_x = - \sin (2 \pi y)/\sqrt{4\pi}$; $\varphi B_y = \sin(4 \pi x)/\sqrt{4\pi}$; $B_z = 0$.
 \end{enumerate}

 \begin{figure}
   \centering
   \subfigure[Mass density.]{
     \includegraphics[width=0.3\linewidth]{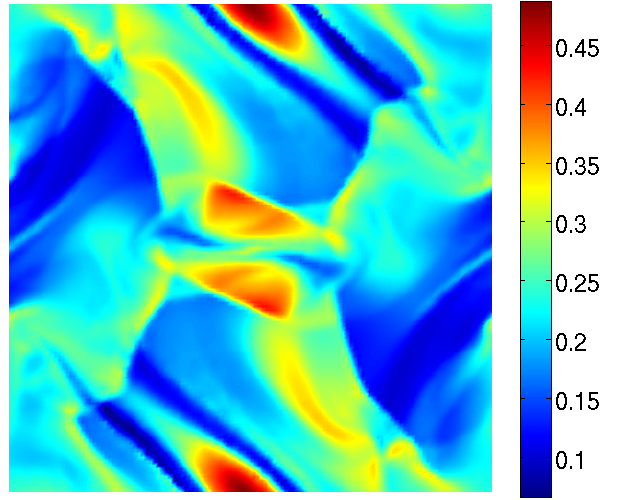}}
   \subfigure[Thermal pressure.]{
     \includegraphics[width=0.3\linewidth]{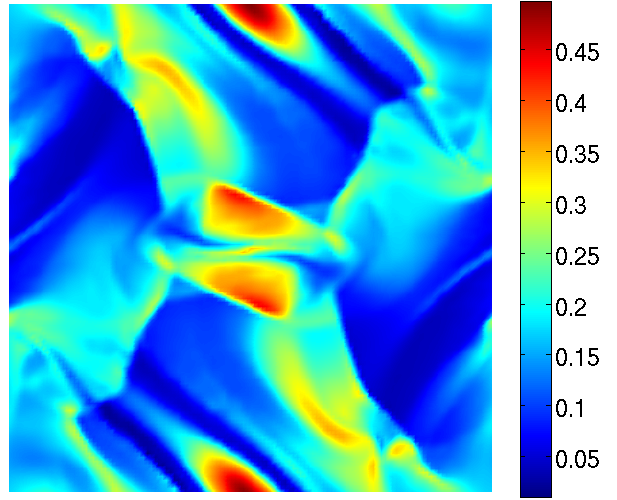}}
   \subfigure[Magnetic pressure.]{
     \includegraphics[width=0.3\linewidth]{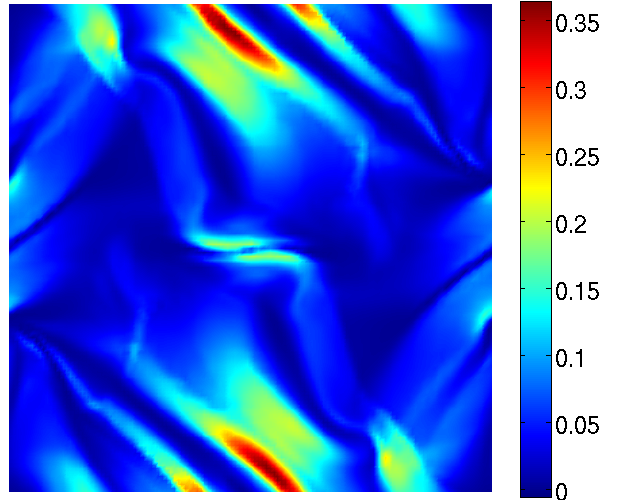}}
   \caption{Results from the Orszag-Tang vortex at time $t = 0.5$, initialised as described in \S\ref{sec:orszag}. }
   \label{fig:orszag-results}
 \end{figure}
 
 The mass density, thermal pressure and magnetic pressure at time $t = 0.5$ are shown in Fig. \ref{fig:orszag-results}, and can be compared with Fig. 22 in \cite{STO08}. The mass density can also be compared with Fig. 16 in \citet{TOT00} or Fig. 14 in \citet{DAI98}. The $\pi$-rotational symmetry of the results implies that our dimensionally-split approach is effective.

\subsection{The rotor problem}
\label{sec:rotor}
Finally, we consider the rotor problem, first proposed in its current form by \citet{BAL99}. This involves placing a rotating disk in the centre of a square domain, which creates rotational discontinuities and propagates Alfv\'en waves into the ambient fluid. The magnetic field is initialised such that it wraps around the rotor, squeezing the fluid from its original circular form. This problem is very useful for testing whether the magnetic field is divergence free, because the build-up of non-zero $\nabla \cdot \vc B = 0$ greatly influences the dynamics. 

We use the initial conditions described in \citet{BAL99}. However, as noted by \citet{TOT00}, the results they present are (erroneously) from a different set of conditions. We stick to what is described in the text, and compare our results with the correct results presented in other papers. We define the radius of the rotor in the centre of the grid as $r < r_0 = 0.1$. This is separated from the ambient fluid by a taper, $r_0 < r < r_1 = 0.115$. We also define $v_0 = 2$, and a constant $f = (r_1 - r)/(r_1 - r_0)$ (which is used for the taper). Then:
 \begin{enumerate}
   \alphen{0}
   \item \textit{Space and time.} Domain: $-0.5 < x, y < 0.5$; number of cells: $256^2$; total time $t = 0.15$.
   \item \textit{Fluid quantities.} The ambient fluid: mass density $\rho = 1$; pressure $p = 1$; velocity $\vc v = 0$; adiabatic index $\gamma = 1.4$. The rotor is defined within a radius $r < 0.1$: mass density $\rho = 10$; $x$-velocity $v_x = -v_0(y - 0.5)/r_0$; $y$-velocity $v_y = v_0 (x - 0.5)/r_0$. The taper is defined within the band $r_0 < r < r_1$: mass density $\rho = 1 + 9f$; $x$-velocity $v_x = -fv_0(y-0.5)/r$; $y$-velocity $v_y = fv_0(x-0.5)/r$. 
   \item \textit{Magnetic field quantities.} Across the entire grid: $B_x = 5$; $B_y = B_z = 0$. 
 \end{enumerate}

 \begin{figure}
   \centering
   \subfigure[Mass density]{
     \includegraphics[width=0.3\linewidth]{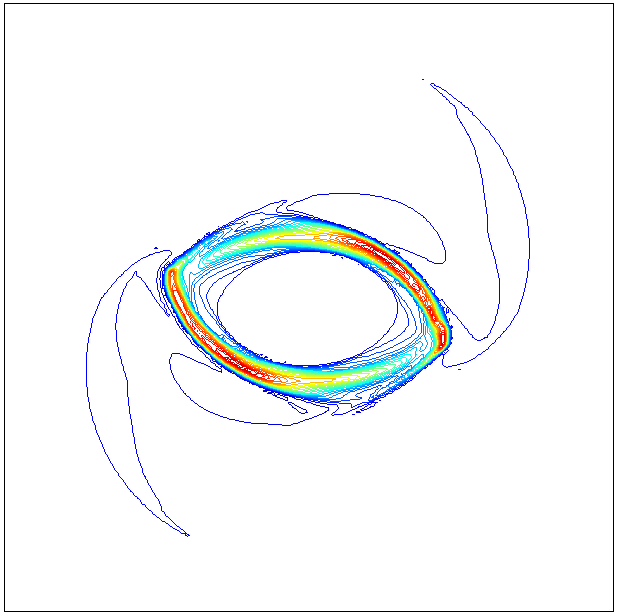}}
   \subfigure[Thermal pressure.]{
     \includegraphics[width=0.3\linewidth]{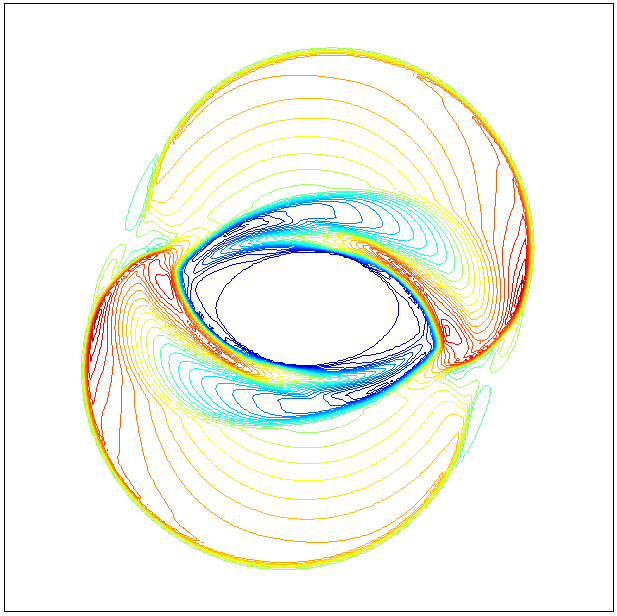}}
   \subfigure[Magnetic pressure.]{
     \includegraphics[width=0.3\linewidth]{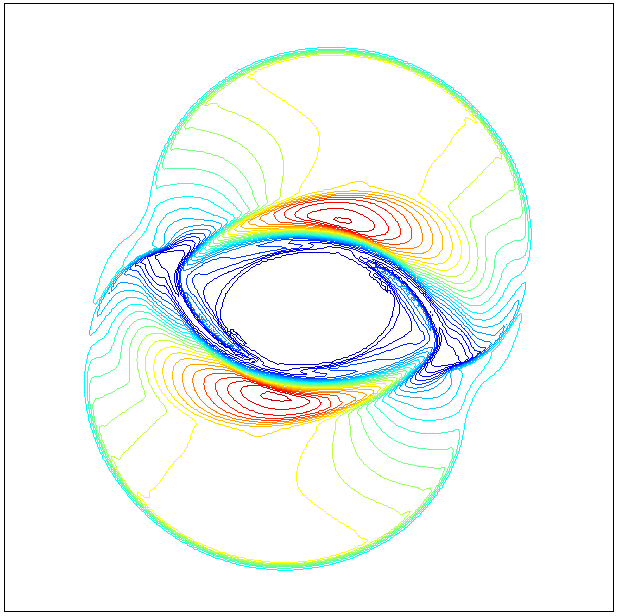}}
   \caption{Results from the rotor problem at time $t = 0.15$, initialised as described in \S\ref{sec:rotor}. All contours are linearly-spaced from 0.426 (blue) to 12.4 (red) for the mass density, 0.0369 (blue) to 1.98 (red) for the thermal pressure and 0.0154 (blue) to 2.60 (red) for the magnetic pressure.}
   \label{fig:rotor-results}
 \end{figure}

The mass density, thermal pressure and magnetic pressure at time $t = 0.15$ are shown in Fig. \ref{fig:rotor-results}. In each plot, there are thirty equally-spaced contours between the highest and lowest values. These results can be compared with Fig. 18 of \citet{TOT00} or Fig. 25 of \citet{STO08}, and imply that the magnetic field is divergence free. Again, it is worth noting the $\pi$-rotational symmetry of the results.

\section{Conclusion}

We have presented a partially dimensionally-split algorithm for numerically solving the equations of MHD. This was done by reformulating the algorithm of \citet{BAL99} and \citet{TOT00}. It is based on the standard reconstruct-solve-average strategy (using a Riemann solver), and relies on constrained transport to ensure the magnetic field remains divergence free ($\nabla \cdot \vc{B} = 0$). The dimensionally-split approach, while not necessarily as accurate as the detailed unsplit algorithm described by \citet{STO08}, comes with many implementational advantages over their approach, and even the Balsara-Spicer-T\'oth algorithm: the information that needs to be stored during the directional passes is minimised; the processing required between complete sets of three directional passes is minimised; the need for staggered grids has been completely eliminated (simplifying the use of adapative and non-uniform grids); the algorithm is more easily parallelisable. All of these advantages also make the addition of this algorithm to existing dimensionally-split codes much simpler. This makes it particularly useful for mature astrophysical codes, which often model more complicated physical effects on top of an underlying hydrodynamics engine and so cannot be restructured easily. We included a complete description of the implementation, and illustrative source code will be made freely available soon. Finally, we demonstrated the accuracy of the algorithm with several common MHD test problems. 

\section*{Acknowledgments}

This research was supported by the Engineering and Physical Sciences Research Council, and has made use of the resources provided by the Edinburgh Compute and Data Facility (ECDF). We would also like to thank Sam Falle at the University of Leeds for his comments.

\footnotesize{
\bibliographystyle{mn2e}
\bibliography{Library/bibliography}}

\label{lastpage}

\end{document}